# Universal turbulence scaling law -8/3 at fusion implosion

Sergei G. Chefranov[*)] and Artem S. Chefranov

Technion-Israel Institute of Technology, Haifa, 32000, Israel

[*)] csergei@technion.ac.il

**Abstract**

The new interpretation of the known results of simulation of the turbulent regime at the time before stagnation stage of the fusion implosion is stated. For this aim the universal turbulence energy spectrum obtained by the authors with a scaling law -8/3, which corresponds to the exact solution of one-dimensional Euler equations for the dynamics of a compressible medium, is used. It is stated that the scaling law -8/3 has more relevance in comparison with the Kolmogorov spectrum of -5/3 in the inertial sub-range of scales for the compressible turbulence at this stage of fusion implosion. A possible mechanism for the occurrence of the anisotropic spectrum -8/3 in turbulence associated with hydrodynamic instability of the rotation of the medium behind the shock wave front is considered.

# 1. Introduction

There is a growing understanding of the degradation sources that appear to have thwarted ignition attempts so far, when obviously one of the reasons is insufficient consideration of the occurrence and development of the strong turbulence in the compression process [1]-[7].

Indeed, the ablation pressure, delivered as a series of weak shocks, accelerates the capsule inwards up to the time when the fuel stagnates (stops moving) and intensive turbulent mixing of the ablator and the D–T fuel can degrade the ability of an implosion to compress the D–T fuel [1], [7]. Imaging data from NIF experiments clearly show the presence of 3-D structures at stagnation and, as it is shown in [2] the extrapolations based on simplified 2-D models differ markedly in their predicted ignition threshold compared to more rigorous 3-D results. However, at an earlier stage of implosion, both 2-D and 1-D modeling of compression modes and their stability can be useful for studying the mechanisms leading to the stagnation stage [6]. At the same time, one of the possible reasons for the instability of implosion and the occurrence of turbulence is the exponentially rapidly developing instability of the motion of the medium behind the converging shock wave front [8],[9]. In [9], the mechanism of instability of a converging cylindrical shock wave is considered, which is associated with the possibility of rotation of the medium behind the shock wave front. Intensification of rotation during implosion, in turn, can lead to a well-known (see, for example, [10]) picture of developed three-dimensional turbulence. The selected direction associated with the axis of rotation can lead to anisotropy of this turbulent regime, similar to the effect of the background magnetic field on the anisotropy of the turbulence spectrum of cosmic plasma [11].

In [11] it is shown that in the perpendicular direction to the vector of the rotation rate inertial wave turbulence has the same, as in the case of kinetic-Alfven waves, physical properties with a spectrum proportional to $k_\perp^{-8/3}$ which propagates towards small scale. Indeed, the anisotropy associated with the background magnetic field, along with the compressibility effect, leads to the one-dimensional turbulence spectrum observed for cosmic plasma, when the Kolmogorov scaling law -5/3 is changed to the -8/3 spectrum for scales smaller than the Larmor radius of ions [11], [12].

It should be noted that the solution to the problem of increasing the efficiency of thermonuclear implosion by taking into account turbulence, which leads to the stagnation of implosion, has so far been largely hindered by the lack of a solution to the problem of a closed statistical description of turbulence itself [12].

Indeed, since the problem of a closed description of single-point and multi-point moments of hydrodynamic fields and the corresponding turbulence spectra remains unsolved, it is also difficult to expect any breakthrough in understanding this problem of increasing the efficiency of thermonuclear implosion.

It should be reminded that, due to the lack of exact explicit solutions to the equations of hydrodynamics, O. Reynolds in 1894 introduced an unclosed infinite system of equations for moments, which led to the up to now unsolved problem of closure. There is only one known exact solution obtained by B. Riemann in 1860 for one-dimensional compressible Euler equations. However, Riemann's solution has an implicit dependence of the solution on the initial conditions and is therefore used only in modern computational fluid dynamics in the form of Riemann solvers.

An explicit solution, as it turned out, can be obtained by applying the theory of generalized functions that were introduced much later by O. Heaviside and P. A. M. Dirac. This made it possible to obtain a closed description of compressible turbulence in [12]-[15], bypassing the problem of closing an infinite chain of Reynolds equations.

For example, in contrast to the theory [11], in [12] for strong turbulence, taking into account interaction between all scales, an exact solution for the turbulent spectrum with an exponent of -8/3 is obtained directly on the basis of an exact explicit analytical solution of one-dimensional Euler hydrodynamic equations in the form of a simple Riemann wave.

Note that for the one-dimensional case, for about forty years there has been used an exact closed statistical description of nonlinear waves described by solutions of partial differential equations of the first order, as for the Riemann wave [16]. However, in [16] and subsequent works [17]-[25], for example, explicit analytical solutions were not obtained for two-point moments of the third and higher order, which left unsolved such problems as the problem of intermittency of turbulence and the problem of dissipative Onsager anomaly [20]. Only the explicit form of the Riemann solution, based on the use of generalized functions, as done in [12], taking into account the theory of [13]-[15], allowed us to give an example of an exact solution to these problems for the case of one-dimensional compressible turbulence. The use of the spectral explicit form of the Riemann solution [22], as well as various modifications of the hodograph method [23]-[25] and methods of symmetry of statistical moments [21] do not provide such an opportunity to obtain a closed analytical description of any single-point and multipoint moments and the corresponding turbulence spectra.

In this paper, the exact solution obtained in [12] for the turbulence energy spectrum corresponding to the scaling law -8/3 is used to interpret the known simulation data [5], [7] of a turbulent regime formed at a relatively early stage of implosion preceding the stagnation stage. In [5] the authors presents results from very high resolution 0.05 $\mu m$ in 3-D implosion simulation of an idealized OMEGA capsule using adaptive mesh refinement Eulerian radiation-hydrocode RAGE are presented. In [7] the detailed description of the behavior of several turbulent quantities at various stage of implosion by using high-resolution 3-D numerical simulation with astrophysical FLASH code is obtained.

The following section 2 shows the exact solution for the universal spectrum -8/3 and compares it in Fig.1 with the simulation data [5] and [7].

Section 3 discusses a possible mechanism for the formation of the -8/3 turbulent spectrum at an early stage of implosion. Analogy in the mechanism of formation of a power exponent -8/3 in the different systems (in a rotating liquid and in the cosmic plasma [11], for the magnetic reconnection [26]-[29] and for the breaking of nonlinear waves [30]) is considered.

## 2. Exact turbulence energy spectrum

To analyze the turbulent flow regime of a compressible medium corresponding to the turbulence energy spectrum of fusion implosion obtained in [5] (see Fig.1), it seems natural to consider the exact solution obtained in [12] for a one-dimensional compressible turbulence energy spectrum (see (63) in [12]):

$$E(k) = C_E k^{-8/3} + O(k^{-10/3}) \qquad (1)$$

$$C_E = \frac{2^{1/3}}{L} \left(\frac{dV_0}{dx}\right)^{8/3}_{x=x_M} \left(\frac{d^3V_0}{dx^3}\right)^{-2/3}_{x=x_M} \Phi^2(0)$$

In (1) $\Phi(z) = \sqrt{\pi} Ai(z)$ -is the Airy function, where $\Phi(0) = \dfrac{\sqrt{\pi}}{3^{2/3}\Gamma(2/3)} \approx 0.629$.

In the turbulence energy spectrum (1), shown in Fig.1 by a black straight line, the initial smooth velocity field $V_0(x)$, $L$ -is the integral external turbulence scale corresponding to this field. To describe the moments and spectra of the turbulent regime in [12], averaging over a spatial variable in an unlimited space is used.

The universality of the spectrum (1) is manifested in the independence of the exponent -8/3 from the type of the initial velocity field. This is due to the fact that (1) is obtained near the collapse of the exact solution of the Euler equations in the limit $t \to t_0$, where $t_0$ is the finite time after which the solution loses smoothness. The time of collapse depends on the initial conditions and, for example, for the case of a poly-tropic medium with an adiabatic exponent $\gamma$, it is defined as (see (33) in [12]) $t_0 = \dfrac{2}{(\gamma+1)\left(\max\limits_{x=x_M}\left|\dfrac{dV_0}{dx}\right|\right)}$.

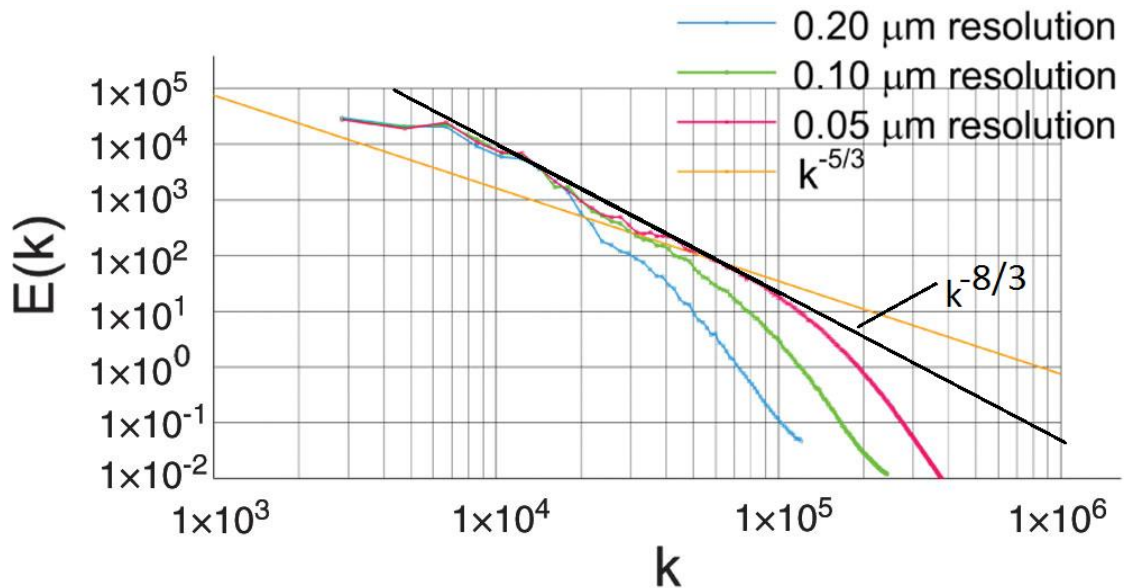

Fig.1 The power spectra of the kinetic energy at t=1.71 ns (at the beginning of stagnation stage in fusion implosion-see Fig.2 below) for three different simulation resolutions demonstrate the emergence of an inertial subrange with the new universal scaling law -8/3 (black line) at the highest possible resolution. Reproduced with permission from Thomas et. al. Phys. Rev. Lett. **109**, 075004

(2012). Copyright 2012 American Physical Society ( Fig. 3 in Ref. [5] ).

RNP/22/Jan/049531.

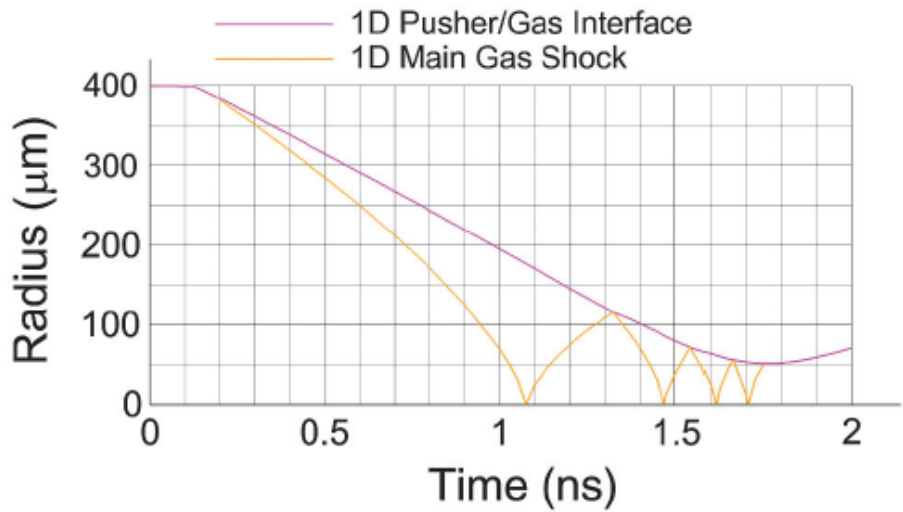

Fig. 2 R-T plot from 1D RAGE simulation of the simplified ICF. Reproduced with permission from Thomas et.al. Phys. Rev. Lett. **109**, 075004 (2012). Copyright 2012 American Physical Society ( Fig. 1a in Ref. [5] ). RNP/22/Jan/049531

Figure 2 shows the dependence of the radius of the converging shock wave until it comes into focus, and then for the reflected shock wave, which is then reflected again sequentially from the piston and from the focus. The turbulent spectrum shown in Fig.1 corresponds to the moment of time near the fourth reflection of the shockwave from the focus. The spectrum in Fig.1, having the scaling law -8/3 in the inertial scale interval, is in accordance with the 1D spectrum obtained also in [7] (see Fig. 12 b [7]) at an earlier stage of implosion, corresponding in Fig.2 to the time of the first reflection of the shockwave from the focus.

## 3. Discussion

A spectrum of turbulence with an exponent of -8/3, as on Fig.1, is obtained in [26] for 1D spectrum of turbulence energy in the case when small-scale magnetic islands or "bubbles" develop near the magnetic reconnection zone. Spectrum -8/3 is obtained also in the numerical simulation for a distribution of bubble sizes generated by the breaking of nonlinear waves [30]. In [26], this spectrum law -8/3 is computed at a time when the energy dissipation rate in the 2-D MHD model is approaching its maximum value while very little (near 5%) of the initial total energy has been dissipated. Indeed, in turbulent plasmas, which contain a large number of small-scale current sheets, reconnection has long been suggested to have a major role in the dissipation of turbulence energy at kinetic scales [27]-[29]. In the turbulence of space plasma in the solar wind and in the magnetosheath the spectrum -8/3 is also observed and is discussed in the connection with the problem of energy dissipation in the space collision-less plasma (see references in [12]). It is shown in [12] that the mechanism responsible for the appearance of a turbulent spectrum in the turbulence of cosmic plasma with a universal scaling index of -8/3 is the collapse of nonlinear Riemann waves occurring in finite time.

In [28], is also observed the violation of the textbook concept of magnetic flux-freezing in the presence of turbulence. In fact, the mechanism of collapse of nonlinear waves in a compressible medium can be the universal mechanism that leads to the magnetic reconnection in the cosmic plasma and to a similar violation of the freezing and invariance of the vortex field, under conditions of quasi-two-dimensional turbulence in a rotating medium. At the same time, instead of magnetic field bubbles, bubbles with a non-zero vortex field will appear, as it turns out as a result of the simulation of the thermonuclear implosion process in [5]. The corresponding process of forming a cascade of vortices of various scales leads, as can be seen from Fig.1, to a universal spectrum of -8/3. This is analogous to the process discussed in [30], in which collapsing nonlinear waves also lead to a -8/3 spectrum for the size distribution for bubbles generated during wave collapse.

Therefore, it is not surprising that there is a good correspondence of spectrum (1) with the results of simulation [5], shown in Fig. 1. In [5], the simulations were designed to both highlight the results of the experiments [4] and address a more general question of how drive asymmetry leads to turbulence generation in the inertial confinement fusion (ICF) implosions. The evidence of fully developed turbulence existing prior to the time of stagnation is significant for ICF applications. It demonstrates how asymmetry can lead to hydrodynamic turbulence by way of instability in a time that is short enough to be of interest for the problem of degrading the compression burn of an ICF implosion [5]. Moreover, the effect of quickly occurring turbulence noted in [5], as shown in [9], can be further enhanced due to the exponentially rapid development of instability of the medium motion behind the converging shock wave front. This instability, obtained in [9] also leads to the rapid emergence of a regime of strong turbulence according to the mechanism established in [10] after the one induced by instability of the intensification to the medium rotation behind the shock front.

Indeed, according to [5], the characteristic value of vorticity $\omega_\varphi \propto 5 \times 10^{11} \sec^{-1}$ in the system before turbulence occurs in it, which allows us to estimate the collapse time as $t_0 \cong \omega_\varphi^{-1} \propto 0.2 \times 10^{-11} \sec$. For this case, the time of formation of a turbulent regime with an energy spectrum (1), according to the estimate given in [12] and made on the base of the method proposed by Onzager, has indeed a very short value $\tau_{8/3} \approx 20 t_0 \propto 0.4 \times 10^{-10} \sec$.

This estimate of the time of formation of the turbulent energy spectrum with an index of -8/3 turns out to be close to the characteristic time $\tau \propto 0.19 \times 10^{-10}$ sec of exponential growth of small perturbations. The later is obtained in [8] for the case of instability of the converging cylindrical shock wave in a gas $\tau \propto \left(\frac{c^2}{\nu}\right)^{-1}$, where $c \approx 888 m/\sec$-is the speed of sound for deuterium for the coefficient of kinematic viscosity $\nu \approx 0.15 cm^2/\sec$.

It seems important that the instability mechanism considered in [9] in the limit of zero viscosity also has a small characteristic time $\tau \propto R/c$ of exponentially rapid instability development determined only by the radius of curvature $R$ of the shock front and the value of the local sound velocity $c$ in the compression region for an arbitrary medium and the equation of state. In [9] the data obtained in simulation and experiment [31] for the converging shock wave in water is used. This mechanism does not depend on the change in the radius of the shock wave over time and may give contribution during all stages of implosion and especially in the period of deceleration of the implosion and stagnation, when the Richtmeyer-Meshkov instability is no longer arising (see [5]). Moreover, the stagnation phase itself can be initiated by such a rapid development of instability of the medium movement behind the shock wave front and the subsequent formation of a turbulent regime.

In the experimental observations and simulations for a converging shock wave in air it was found that during implosion, vortex structures also appear in the medium behind the shock wave front [32]-[35].

Thus, as can be seen from Fig.1, the use of the scaling law -8/3 from (1) seems to be more adequate than scaling law -5/3 represented in [5] for the moment of time 1.71 ns, corresponding to the stagnation of implosion (see Fig.2). In [7], for the same stage of implosion, the turbulence energy spectrum is also given, which is compared with the law -5/3, that is realized only at the stagnation stage of implosion. Also it turns out that the law -8/3 in a wider range of wavenumbers corresponds to the simulation data [7] than the law -5/3 at all the stages of implosion in a wider range of scales. Moreover, at the previous stages, starting from the moment of the first collapse of the shock in focus, the universal scaling law -8/3 also corresponds well to the obtained spectra in [7] within the same limits of the change in the wavenumbers. Opposite to it, the scaling law -5/3 is not arising during earlier stages of fusion implosions [7].

Thus, the implementation of the law -8/3 of (1), corresponding to simulations [5] and especially [7], yet at the stage of accelerated implosion, indicates the existence of a new mechanism for the occurrence of turbulence, which was not previously taken into account. The above-mentioned rapid development of the instability considered in [7], may precede the implementation of such a mechanism of turbulence in a compressible medium. Such a new mechanism can be associated with the collapse of a simple nonlinear Riemann wave in a finite time, which is similar to the mechanism of formation of the universal -8/3 turbulence spectrum, which is observed for the turbulence spectrum in the space plasma of the magnetosphere and the solar wind (see Fig.1 and Fig. 2 in [12]).

Most significantly, imaging data from NIF experiments clearly show the presence of 3-D structures at stagnation of the fusion implosion [1]-[3]. The question arises about the reason for the coincidence of the one-dimensional spectrum (1) indicated in Fig. 1 with the data of the simulation of three-dimensional turbulence and its spectrum during the fusion implosion.

A possible reason for this is a violation of spherical symmetry during the fusion implosion due to the occurrence of rotation of the medium behind the front of a converging spherical shock wave by the instability mechanism similar to the noted for converging cylindrical shock waves [8], [9]. In this case, a dedicated direction is created and an adequate description of an anisotropic three-dimensional turbulence is provided by using a universal one-dimensional spectrum -8/3 of (1). Indeed, in [11], on a basis of a numerical simulation, it is stated that in the perpendicular direction to the direction of rotation, the inertial waves turbulence has an anisotropic one-dimensional scaling law–8/3 as in the anisotropic turbulence spectrum in the space plasma.

## 4. Conclusions

The use of the new theory of turbulence allowed us to give a new interpretation to the known results of simulation of the turbulent regime that occurs during the fusion implosion and limits the effectiveness of this process. It facilitates the choice of direction for the experimental research to provide an appropriate confirmation of the mechanism of turbulence development considered above at the stage of accelerated implosion. Thus, a new understanding of the mechanisms by which a turbulent regime arises at early stages of a fusion implosion can enable the development of methods that increase the efficiency of energy output in this process.

We thank A. G. Chefranov and Ya. E. Krasik.

The study is support by the Israel Science Foundation, Grant No. 492/18.

## Data Availability Statement

The data that support the findings of this study are available from the corresponding author upon reasonable request.

## Literature

1. O. A. Hurricane , D. A. Callahan, D. T. Casey, et.al., Fuel gain exceeding unity in an inertially confined fusion implosion, Nature (London), **506**, 343 (2014); https://doi.org/10.1038/nature13008

2. D. S. Clark, C. R. Weber, J. L. Milovich , et.al., Three-dimensional modeling and hydrodynamic scaling of National Ignition Facility, Phys. Plasmas, **26**, 050601 (2019) ; https://doi.org/10.1063/1.5091449

3. D. J. Schlossberg, G. P. Gim, D. T. Casey, et. al., Observation of hydrodynamic flows in imploding fusion plasmas of National Ignition Facility, Phys. Rev. Lett., **127**, 125001 (2021); https://doi.org/10.1103/PhysRevLett.127.125001

4. J. R. Rygg , J. A. Frenje , C. K. Li ,et.al., Observations of the collapse of asymmetrically driven convergent shocks,  Phys. Plasmas, **15,** 034505 (2008); https://doi.org/10.1063/1.2892025

5. V. A. Thomas, and R. J. Kares, Drive asymmetry and the origin of turbulence in an ICF implosion, Phys. Rev. Lett., **109**, 075004 (2012); https://doi.org/10.1103/PhysRevLett.109.075004

6. R. S. Craxton, K. S. Anderson, T. R. Boehly, et. al., Direct-drive inertial confinement fusion: A review, Phys. Plasmas **22**, 110501 (2015); http://dx.doi.org/10.1063/1.4934714

7. I. Boureima, P. Ramaprabhu, and N. Attal, Properties of the turbulent mixing layer in a spherical implosion, J. Fluid Eng., **140**, 050905 (2018); https://doi.org/10.1115/1.4038401

8. S. G. Chefranov, Dissipative instability of converging cylindrical shock wave, Phys. Fluids, **32**, 114103 (2020); https://doi.org/10.1063/5.0027109

9. S. G. Cherfanov, Instability of cumulation in converging cylindrical shock wave, Phys. Fluids, **33**, 096111 (2021); https://doi.org/10.1063/5.0065017

10. J. P. Gollub, and H. L. Swinney, Onset of turbulence in a rotating fluid, Phys. Rev. Lett., **35**, 927 (1975)

11. S. Galtier, and V. David, Inertial/kinetic-Alfven wave turbulence: A twin problem in the limit of local interactions, Phys. Rev. Fluids, **5**, 044603 (2020); https://doi.org/10.1103/PhysRevFluids.5.044603

12. S. G. Chefranov, and A. S. Chefranov, Exact solution to the main turbulence problem for a compressible medium and the universal -8/3 law turbulence spectrum of breaking waves Phys. Fluids, **33**, 076108 (2021); https://doi.org/10.1063/5.0056291

13. S. G. Chefranov, Exact statistically closure description of vortex turbulence and admixture in compressible medium. Sov. Phys. Dokl. , **36** 286(1991)

14. S. G. Chefranov, and A. S. Chefranov , Exact solution of the compressible Euler–Helmholtz equation and the millennium prize problem generalization. Phys. Scr., **94**, 054001 (2019)

    https://doi.org/10.1088/1402-4896/aaf918

15. S. G. Chefranov, and A. S. Chefranov , The new exact solution of the compressible 3D Navier–Stokes equations. Commun. Nonlinear Sci. Numer. Simul., **83**,105118 (2020)

    https://doi.org/10.1016/j.cnsns.2019.105118

16. A. N. Malakhov, and A. I. Saichev, On the question of kinetic equations in the theory of random waves. Proc. of Higher Edu. Instit., Radiophysics, **17**, 699 (1974)

17. S. N. Gurbatov, A. I. Saichev, and A. N. Malakhov, Nonlinear random waves and turbulence in nondispersive media: waves, rays, particles Manchester University Press, 1991, and John Wiley and Sons Ltd, 1992, 308 p

18. E. A. Novikov, A new approach to the problem of turbulence, based on the conditionally averaged Navier-Stokes equations, Fluid Dyn. Res., **12**, 107 (1993)

19. A. M. Polyakov, Turbulence without pressure. Phys. Rev. E, **52**, 6183 (1995)

20. K. R. Sreenivasan, and V. Yakhot, Dynamics of three-dimensional turbulence from Navier-Stokes equations. Phys. Rev. Fluids., **6**, 104604 (2021); https://doi.org/10.1103/PhysRevFluids.6.104604

21. M. Oberlack, S. Hoyas, S. V. Kraheberger, F. Alcantara-Avila, and J. Laux, Turbulence statistics of arbitrary moments of wall-bounded shear flows: A symmetry approach, Phys. Rev. Lett., **128**, 024502 (2022)

22. E. N. Pelinovskii, Spectral analysis of simple waves, Radiophys. Quantum. Electron., **19**, 262 (1976) http://dx.doi.org/10.1007/BF01034583

23. D. B. Fairlie, Equations of hydrodynamic type. www.arxiv.org/hep-th/9305049v1  12 May 1993

24. D. B. Fairlie, Integrable systems in higher dimensions. Prog. Theor. Phys. Suppl. , **118**, 309 (1995); https://doi.org/10.1143/PTPS.118.309

25. B. G. Konopelchenko, and G. Ortenzi, On universality of homogeneous Euler equation. J. Phys. A: Math. Theor. , **54**, 205701 (2021)

    https://doi.org/10.1088/1751-8121/abf586

26. W. H. Matthaeus, and S. L. Lamkin, Turbulent magnetic reconnection, Phys. Fluids, **29**, 2513 (1986); https://doi.org/10.1063/1.866004

27. T. D. Phan, et.al., Electron magnetic reconnection without ion coupling in Earth's turbulent magnetosheath, Nature, **557**, 202 (2018);

    https://doi.org/10.1038/S41586-018-0091-5

28. A. Lazarian, G. L. Eyink, A. Jafari, G. Kowal, H. Li, S. Xu, and E. T. Vishniac,3D turbulent reconnection: Theory, tests, and astrophysical implications, Phys. Plasmas, **27**, 012305 (2020); https://doi.org/10.1063/1.5110603

29. P. Shi, P. Srivastov, M. H. Barbhuiya, P. A. Cassak, E. E. Scime, and M. Swisdak, Laboratory observations of electron heating and non-Maxwellian  distributions of the kinetic scale during electron-only magnetic reconnection, Phys. Rev. Lett., **128**, 025002 (2022); https://doi.org/10.1103/PhysRevLett.128.025002

30. W. H. R. Chan, P. L. Johnson, P. Moin, and J. Urzay, The turbulent bubble break-up cascade. Part 2. Numerical simulation of breaking waves, J. Fluid Mech. **912**, A43 (2021); https://doi.org/10.1017/jfm.2020.1084

31. A. Rososhek, D. Nozman, and Ya. E. Krasik, Addressing the symmetry of a converging cylindrical shock wave in water close to implosion, Appl. Phys. Lett., **118**, 174103 (2021); https://doi.org/10.1063/5.0050033
32. S. K. Takayama, H. Klein, and H. Gronig, An experimental investigation of the stability of converging cylindrical shock waves in air, Exp. Fluids **5**, 315 (1987)
33. M. Watanabe and K. Takayama, Stability of converging cylindrical shock waves, Shock Waves **1**, 149 (1991)
34. M. Watanabe and K. Takayama, Stability of converging cylindrical shock waves, JSME Int. J. Ser. II **35**, 218 (1992)
35. M. Kjellander, "Energy concentration by converging shock waves in gases," Technical Report No. SE-100-44 (Royal Institute of Technology (KTH) Mechanics, SE-100 44 Stockholm, Sweden, 2012), p. 90.